\documentclass[useAMS,usenatbib]{mn2e}
\usepackage{graphicx}
\usepackage{natbib}
\usepackage[T1]{fontenc}
\usepackage{aecompl}
\usepackage{color}
\newcommand{\comments}[1]{}

\def\mv{$m_{\mathrm{V}}~$}

\def\ugr{u$^{\prime}$g$^{\prime}$r$^{\prime}$}
\def\ugi{u$^{\prime}$g$^{\prime}$i$^{\prime}$}

\title[Testing the planetary models of HU~Aquarii]
	{Testing the planetary models of HU~Aquarii}
\author[M.C.P. Bours et al.]
  {M.C.P.~Bours$^1$\thanks{m.c.p.bours@warwick.ac.uk},
  T.R.~Marsh$^1$,
  E.~Breedt$^1$, 
  C.M.~Copperwheat$^2$,
  V.S.~Dhillon$^3$,
  \newauthor 
  A.~Leckngam$^4$,
  S.P.~Littlefair$^3$,
  S.G.~Parsons$^5$,
  A.~Prasit$^4$. \\
  $^1$Department of Physics, University of Warwick, Coventry CV4 7AL, UK. \\
  $^2$Astrophysics Research Institute, Liverpool John Moores University, Twelve Quays House, Birkenhead, CH41 1LD, UK. \\
  $^3$Department of Physics and Astronomy, University of Sheffield, Sheffield, S3 7RH, UK. \\
  $^4$National Astronomical Research Institute of Thailand, 191 Siripphanich Building, Huay Kaew Road, Chiang Mai 50200, Thailand. \\
  $^5$Departmento de F\'isico y Astronom\'ia, Universidad de Valpara\'iso, Avenida Gran Bretana 1111, Valpara\'iso, Chile.}

\date{Accepted XX XXXX -- Received XX XXXX}

\pagerange{\pageref{firstpage}--\pageref{lastpage}} \pubyear{2002}

\def\LaTeX{L\kern-.36em\raise.3ex\hbox{a}\kern-.15em
    T\kern-.1667em\lower.7ex\hbox{E}\kern-.125emX}

\begin{document}

\label{firstpage}

\maketitle

\begin{abstract}
We present new eclipse observations of the polar (i.e. semi-detached magnetic white dwarf + M-dwarf binary) HU~Aqr, and mid-egress times for each eclipse, which continue to be observed increasingly early. Recent eclipses occurred more than 70~seconds earlier than the prediction from the latest model that invoked a single circumbinary planet to explain the observed orbital period variations, thereby conclusively proving this model to be incorrect. Using ULTRACAM data, we show that mid-egress times determined for simultaneous data taken at different wavelengths agree with each other. The large variations in the observed eclipse times cannot be explained by planetary models containing up to three planets, because of poor fits to the data as well as orbital instability on short time scales. The peak-to-peak amplitude of the O-C diagram of almost 140~seconds is also too great to be caused by Applegate's mechanism, movement of the accretion spot on the surface of the white dwarf, or by asynchronous rotation of the white dwarf. What does cause the observed eclipse time variations remains a mystery.

\end{abstract}


\section{Introduction}
HU~Aqr was discovered independently by \citet{Schwope93} and \citet{Hakala93} as an eclipsing binary of the AM~Her type, also known as polars. This subset of cataclysmic variables (CVs) contains a Roche-lobe filling M-dwarf secondary star and a strongly magnetic ($\sim$~10~MG) white dwarf as the primary star. From its discovery onwards, HU~Aqr has been studied extensively and over a wide range of wavelengths \citep[e.g.][]{Glenn94, Schwope97, Schwope01, Heerlein99, Howell02, HarropAllin01, HarropAllin99, Vrielmann01, Bridge02, Watson03, Schwarz09}. The general picture is as follows: the M-dwarf loses matter at the L1 Lagrange point, which then follows a ballistic trajectory until the ram pressure equals the white dwarf's magnetic pressure and the stream couples to the magnetic field lines. At that point the stream leaves the orbital plane and is guided along the field lines until it accretes onto the white dwarf's magnetic pole, creating a luminous accretion spot. 

The accretion rate in this system is highly variable, and the binary has changed from a high to a low state and back several times over the last decades. The variability causes both flickering typical of CVs on timescales of minutes, and significant changes in the overall shape of the observed light curves on timescales as short as one orbital cycle \citep{HarropAllin01}. 

One constant in the light curves is the eclipse of the white dwarf by the secondary star, the ingress and egress of which last for $\sim$~30~seconds. In a high state the eclipse is dominated by contributions from the accretion spot and stream, while the ingress and egress of the white dwarf itself is hardly visible. Due to the geometry of the system the accretion spot and accretion stream are well separated during egress, giving this part of the light curve a fairly constant shape. During ingress the distinction between the accretion spot and stream features is less clear. From X-ray data there is also evidence of enhanced absorption by an accretion curtain along the ballistic stream at this phase \citep{Schwope01}. To accurately determine the eclipse times the timing has therefore been based on the egress of the accretion spot. Comparison of observed mid-egress times to expected mid-egress times, which are calculated assuming a constant orbital period, have revealed considerable variations \citep{Schwope14,Gozdziewski12,Schwarz09,Schwope01}. 

Several explanations for these eclipse time variations have been offered in the literature. Similar variations have been seen in a number of other binaries, leading \citet{Applegate92} to propose a theory that has since become known as Applegate's mechanism. It assumes that the M-dwarf experiences Solar-like magnetic cycles that couple to the binary's orbital period, by affecting the gravitational quadrupole moment of the M-dwarf, and therefore cause genuine modulations of the orbital period. However, for HU~Aqr, the energy available in the M-dwarf is insufficient to explain the large variations observed \citep{Schwarz09}. A second explanation for the eclipse time variations is that they are caused by the presence of planet-like or brown dwarf-like bodies in wide orbits around the binary. The additional mass causes periodic shifts in the binary's centre of mass, which are reflected in the eclipse times \citep{Marsh14}. This theory has gained popularity after the discovery of circumbinary planets around double main-sequence star binaries \citep[e.g.][]{Welsh12,Orosz12a}, and models with 1, 2 and even 3 planets have been proposed for HU~Aqr \citep{Gozdziewski12,Qian11}. All have since been disproved on grounds of dynamical instability \citep{Wittenmyer12,Horner11} or by new data \citep[this paper,][]{Schwope14,Gozdziewski12}. 

In this paper we present 22 new eclipse times from data taken between June 2010 and June 2014.

\section{Observations}
In this section we describe the technical details of the observations. The light curves themselves are discussed in Sect.~\ref{sect:ever_changing_light_curves}. All mid-egress times are listed in Table~\ref{tab:times}, which also summarises the technical details and contains notes on the observing conditions. Details of how we determined the mid-egress times are described in Sect.~\ref{sect:egresstimes}.

\subsection{RISE on the Liverpool Telescope}
\begin{figure}
\begin{center}
\includegraphics[]{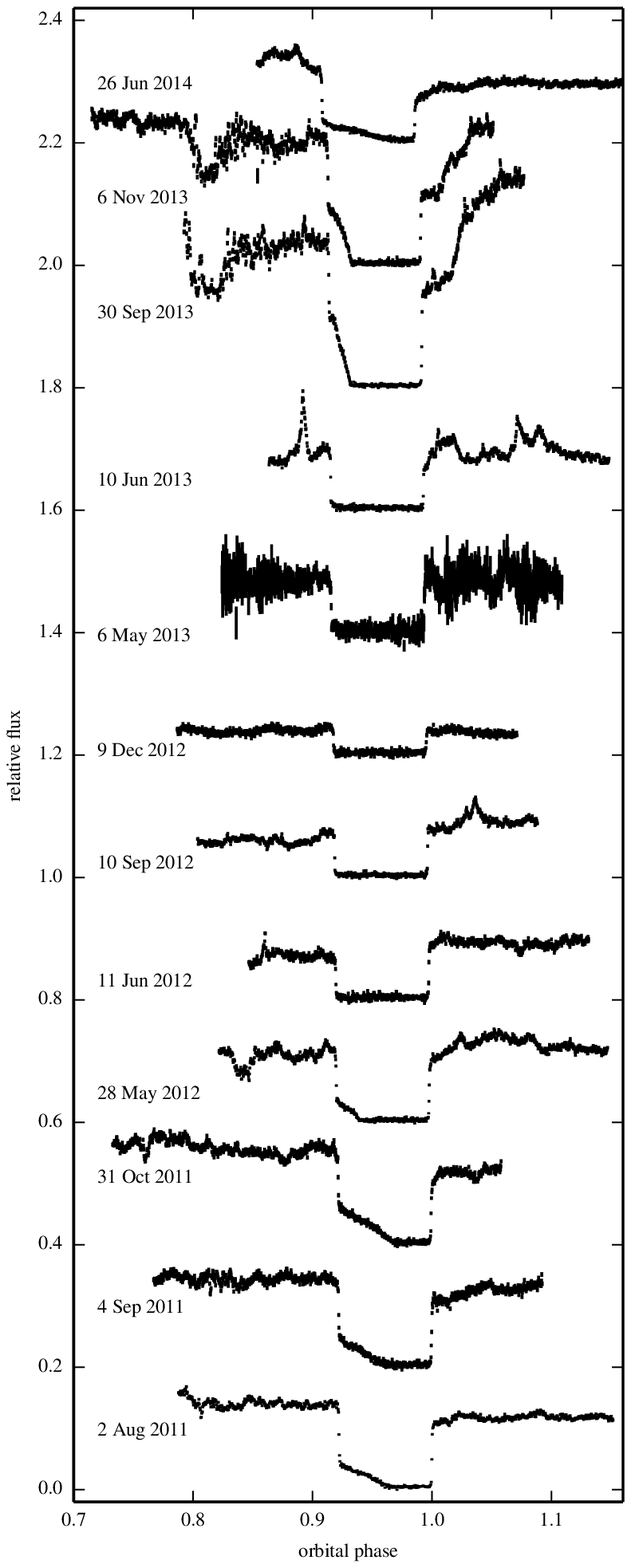}
\caption{LT+RISE light curves of HU~Aqr, taken between 2 Aug 2011 and 26 Jun 2014. The vertical axis shows the flux relative to the comparison star, which is the same star for all LT+RISE and TNT+ULTRASPEC data. Each light curve is vertically offset from the previous one by 0.2. To calculate the orbital phase we used the ephemeris in equation~\ref{eq:ephemeris}.} 
\label{fig:lteclipses}
\end{center}
\end{figure}

We observed 12 eclipse observations of HU~Aqr between 2 Aug 2011 and 26 Jun 2014 with the Liverpool Telescope \citep[LT;][]{Steele04}, a 2-metre robotic telescope on the island of La Palma, Spain. The instrument used was the RISE camera \citep{Steele08}, which contains a frame transfer CCD and uses a single `V+R' filter, and we used the 2x2 binned mode. The data were flatfielded and debiased in the automatic pipeline, in which a scaled dark-frame was removed as well. Aperture photometry was then performed using the ULTRACAM pipeline \citep{Dhillon07}, and care was taken to use the same comparison star for all data reductions. This comparison is a non-variable star located 93" South and 88" West of HU~Aqr. We chose this star rather than comparison `C' as in \citet[][their Fig.~1]{Schwope93} because of their relative brightnesses in RISE's V+R filter. We did use comparison `C' to calculate magnitudes for HU~Aqr, as will be explained in Sect.~\ref{sect:ever_changing_light_curves}. All stellar profiles were fitted with a Moffat profile \citep{Moffat69}, the target aperture diameters scaled with the seeing and the comparison star was used to account for variations in the transmission. The light curves are shown in Fig.~\ref{fig:lteclipses}.

\subsection{ULTRACAM observations}
\begin{figure*}
\begin{center}
\includegraphics[width=\textwidth]{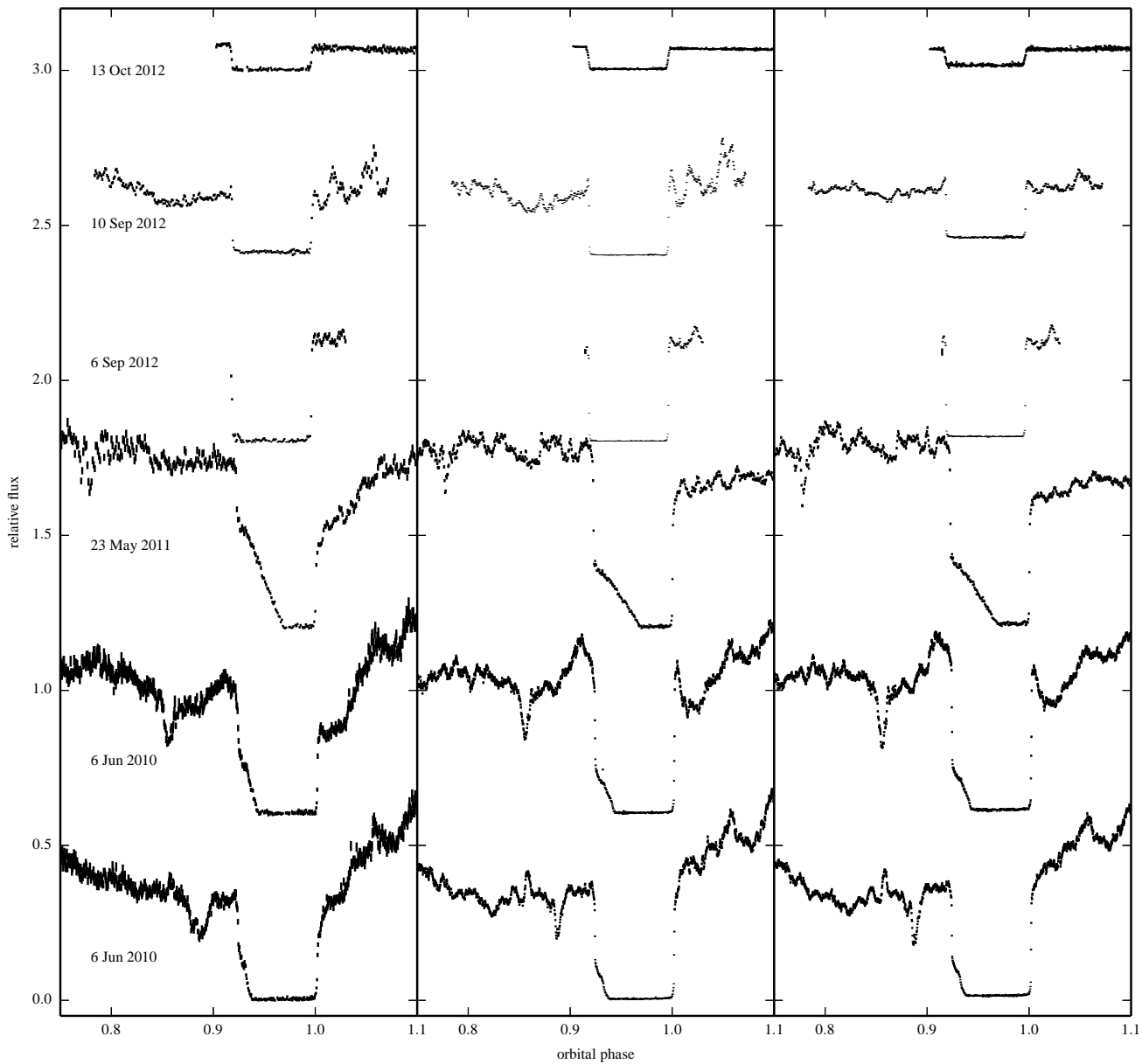}
\caption{ULTRACAM+NTT (bottom three) and ULTRACAM+WHT (upper three) light curves of HU~Aqr, taken between 6 Jun 2010 and 13 Oct 2012. From left to right are shown the blue (u$^{\prime}$), green (g$^{\prime}$), and red (r$^{\prime}$ or i$^{\prime}$) data. The blue and red light curves are scaled to the green light curves to facilitate easy comparison. Each light curve is vertically offset from the previous one by 0.6. To calculate the orbital phase we used the ephemeris from equation~\ref{eq:ephemeris}. }
\label{fig:ultracameclipses}
\end{center}
\end{figure*}

Between June 2010 and October 2012 we obtained six eclipses of HU~Aqr with the high-speed camera ULTRACAM, which splits the incoming light into three beams, each containing a different filter and a frame-transfer CCD \citep{Dhillon07}. ULTRACAM was mounted on the New Technology Telescope (NTT, 3.5m) in Chile during the first three observations and on the William Herschel Telescope (WHT, 4.2m) on La Palma, Spain, for the last three observations. The light curves are shown in Fig.~\ref{fig:ultracameclipses}.

For the data reduction and the relative aperture photometry we used the ULTRACAM pipeline. Due to different fields of view and windowed setups during the observations we could not use the same comparison star as for the LT+RISE and TNT+ULTRASPEC data. 

\subsection{ULTRASPEC on the Thai National Telescope}
In November 2013 we observed two HU~Aqr eclipses with the 2.4-metre Thai National Telescope (TNT), located on Doi Inthanon in northern Thailand. We used the ULTRASPEC camera \citep{Dhillon14}, which has a frame-transfer EMCCD, with an SDSS g$^{\prime}$ filter on November 10, and a Schott KG5 filter on November 13. The Schott KG5 filter is a broad filter with its central wavelength at 5075~\AA~and a FWHM of 3605~\AA. 

\begin{figure}
\begin{center}
\includegraphics[]{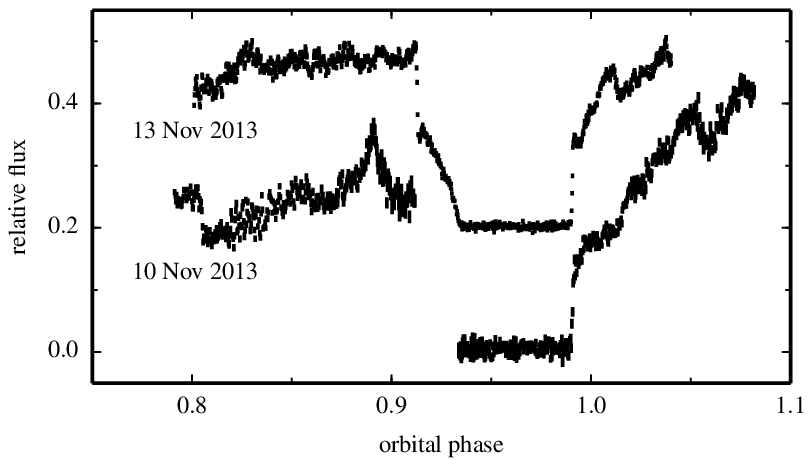}
\caption{TNT+ULTRASPEC light curves of HU~Aqr, taken at 10 Nov 2013 (bottom) and 13 Nov 2013 (top, vertically offset by 0.2). To calculate the orbital phase we used the ephemeris from equation~\ref{eq:ephemeris}. }
\label{fig:tnt_eclipses}
\end{center}
\end{figure}

The data were reduced using the ULTRACAM pipeline, with which we debiased and flatfielded the data and performed aperture photometry. We used the same comparison star as for the reduction of the LT+RISE data. Both TNT light curves are shown in Fig.~\ref{fig:tnt_eclipses}. During the observations on November 10, the telescope briefly stopped tracking, causing the gap seen during ingress in the light curve in Fig.~\ref{fig:tnt_eclipses}. 

\subsection{Wide Field Camera on the Isaac Newton Telescope}
In June 2014 we observed two HU~Aqr eclipses using the Wide Field Camera (WFC) mounted at the prime focus of the 2.5-metre Isaac Newton Telescope (INT) on La Palma, Spain. The read-out time of the WHF is $\sim$~2~seconds, and we used a Sloan-Gunn g filter.

\begin{figure}
\begin{center}
\includegraphics[]{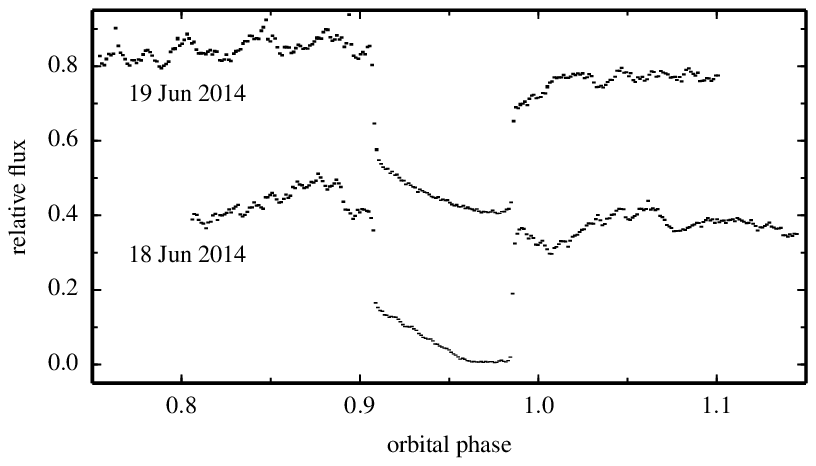}
\caption{INT+WFC light curves of HU~Aqr, taken at 18 Jun 2014 (bottom) and 19 Jun 2014 (top, vertically offset by 0.4). To calculate the orbital phase we used the ephemeris from equation~\ref{eq:ephemeris}. }
\label{fig:intwfc_eclipses}
\end{center}
\end{figure}

The data were reduced using the ULTRACAM pipeline, with which we debiased and flatfielded the data and performed aperture photometry. As a comparison star we used star `C' as in \citet{Schwope93}. Both light curves are shown in Fig~\ref{fig:intwfc_eclipses}.

\section{Ever-changing light curves} \label{sect:ever_changing_light_curves}
HU~Aqr is known for its variable accretion rate, which significantly influences the brightness of the system and the morphology of its light curves, as is immediately clear from Figs.~\ref{fig:lteclipses}, \ref{fig:ultracameclipses}, \ref{fig:tnt_eclipses} and  \ref{fig:intwfc_eclipses}. HU~Aqr was in a high state until mid 2012, then went into a low state, and returned to a high state in the second half of 2013, after which it dropped to a lower state again. We calculated the magnitude of HU~Aqr for each dataset, using the out-of-eclipse data, and excluding obvious flares from the M-dwarf and dips due to the accretion stream. Using simultaneous g$^{\prime}$ and r$^{\prime}$ ULTRACAM data, we determined a magnitude offset for comparison `C' \citep[][their Fig.~1]{Schwope93} for each filter type with respect to its V-band magnitude of \mv = 14.65 (mean of various measurements). We then calculated magnitudes for HU~Aqr as normal, using the relative flux and corrected magnitude for comparison `C'. These approximate magnitudes are listed in Table~\ref{tab:times}. Typical magnitudes were $m_{g^{\prime}}$ $\simeq$ 15.1 - 15.5 during the high states, and $m_{g^{\prime}}$ $\simeq$ 17.8 during the low state. 

The out-of-eclipse data shows the characteristic flickering of a CV, caused by the irregular, blobby nature of the accretion. In these high state light curves the eclipse is dominated by both the accretion spot on the white dwarf and the accretion stream. The eclipse of the spot is characterised by the abrupt ingress when the spot goes behind the M-dwarf, and the egress when it re-emerges, each typically lasting $\sim$~8~seconds. 

Due to the small physical size of the accretion region on the white dwarf, this spot can reach very high temperatures and contributes significantly to the total light from the system when accretion rates are high. Our data shows no signs of a varying width or height of the ingress or egress feature, over time nor with changing accretion rate. Assuming for the masses $M_{\mathrm{wd}}$ = 0.80~M$_{\odot}$ and $M_2$ = 0.18~M$_{\odot}$, an inclination $i$ = 87$^{\circ}$ \citep{Schwope11} and the orbital period from equation~\ref{eq:ephemeris}, we arrive at an orbital velocity of the M-dwarf of 479~km/s and a maximum spot diameter of $D_{\mathrm{spot}} \simeq$ 3829~km. Using the mass-radius relation of \citet{VR88} for the white dwarf, this corresponds to a fractional area on the white dwarf of 0.018 and an opening angle of the accretion spot of $\sim 30^{\circ}$. This is significantly larger than the value of 3$^{\circ}$ found by \citet{Schwope01} using ROSAT-PSPC soft X-ray data although some difference is to be expected, as only the hottest parts of the spot will radiate at X-ray wavelengths.

In the high state the ingress of the accretion stream, caused by irradiation of the stream by the hot accretion spot, is visible as an additional, shallower slope after the sharp ingress of the accretion spot. Depending on the exact geometry of the stream and its relative position to the spot, the duration and height of the stream ingress vary considerably. 

Some of the light curves taken during a high state also show a narrow dip near orbital phase 0.8-0.9. This has been seen in many of the other studies on HU~Aqr, and is also observed in other polars \citep{Watson95}. It is caused by the obscuration of the accretion spot by part of the stream if the inclination of the system is such that the hemisphere with both the spot and stream is towards the observer \citep{Bridge02}. The depth and width of this dip carry information about the temperature difference between the white dwarf and the spot and about the physical extent of the magnetically coupled stream respectively. The dip moves further away from the spot ingress during high accretion states, which agrees with the expectation that the ballistic stream penetrates further before coupling to the magnetic field lines in the high state. We also notice some interesting colour differences in the dip as well as in some of the other variable features in the high state ULTRACAM light curves (Fig.~\ref{fig:ultracameclipses}). At blue wavelengths the dip is wider than observed in the other two bands, and it possibly consists of two components. This indicates a fluffy and blobby nature of the stream, and a strongly wavelength-dependent opacity within the stream.

During a low state, the stream ingress disappears completely, flickering is less pronounced and the system is noticeably fainter.

\section{Egress times} \label{sect:egresstimes}
\begin{table*}
\begin{center}
\caption{Mid-egress times for the observed eclipses of HU~Aqr. For the ULTRACAM data the exposure times are listed for each of the arms separately and the mid-egress times are the weighted averages of the three times from the individual arms. The magnitudes are calculated as detailed in the text and correspond to the given filter of each observation and the g$^{\prime}$-filter for the ULTRACAM data. Dead-time between each exposure is 8~ms (LT+RISE), 25~ms (ULTRACAM), 15~ms (ULTRASPEC) and 2~s (INT+WFC).}
\label{tab:times}
\begin{tabular}{l l l l l l l l}
\hline \hline
date    & cycle  & mid-egress time & t$_{\mathrm{exp}}$ & telescope + & filter(s) & approximate & observing conditions  \\
        & number & BMJD(TDB)       & (sec)              & instrument  &           & magnitude   &              \\
\hline \hline
06 Jun 2010 & 72009 & 55354.2706040(4)  & 4, 2, 2 & NTT+UCAM  & \ugr         & 15.8 & clear, seeing 1.5\arcsec           \\
06 Jun 2010 & 72010 & 55354.3574451(5)  & 4, 2, 2 & NTT+UCAM  & \ugr         & 15.7 & clear, seeing 1-2\arcsec           \\
23 May 2011 & 76053 & 55705.3721585(9)  & 8, 4, 4 & NTT+UCAM  & \ugr         & 15.5 & clear, seeing 1.5\arcsec           \\
02 Aug 2011 & 76868 & 55776.1307426(20) & 3       & LT+RISE   & V+R          & 15.9 & clear, seeing 2\arcsec   \\
04 Sep 2011 & 77247 & 55809.0356564(25) & 2       & LT+RISE   & V+R          & 15.8 & clear, seeing 2.5\arcsec \\
31 Oct 2011 & 77902 & 55865.9029864(22) & 2       & LT+RISE   & V+R          & 15.7 & clear, seeing 1.6-2.5\arcsec \\
28 May 2012 & 80324 & 56076.1818394(22) & 2       & LT+RISE   & V+R          & 15.9 & clear, seeing 1.8-2.4\arcsec \\
11 Jun 2012 & 80485 & 56090.1598976(19) & 2       & LT+RISE   & V+R          & 17.3 & thin clouds, seeing 2-4\arcsec \\
06 Sep 2012 & 81486 & 56177.0670248(8)  & 7, 4, 4 & WHT+UCAM  & \ugr         & 16.1 & clear, seeing 1-2\arcsec         \\
10 Sep 2012 & 81531 & 56180.9739470(22) & 2       & LT+RISE   & V+R          & 16.4 & thin clouds, seeing 2.2\arcsec \\
10 Sep 2012 & 81532 & 56191.0607721(6)  & 7, 4, 4 & WHT+UCAM  & \ugi         & 16.6 & thin clouds, seeing 2\arcsec \\
13 Oct 2012 & 81910 & 56213.8788462(2)  & 6, 2, 2 & WHT+UCAM  & \ugr         & 17.8 & thin clouds, seeing 2-4\arcsec     \\
09 Dec 2012 & 82566 & 56270.8329602(53) & 2       & LT+RISE   & V+R          & 17.2 & clear, seeing 2\arcsec   \\
06 May 2013 & 84275 & 56419.208883(11)  & 2       & LT+RISE   & V+R          & 17.3 & thick clouds, seeing 2-3\arcsec \\
10 Jun 2013 & 84678 & 56454.1974374(17) & 2       & LT+RISE   & V+R          & 16.2 & clear, seeing 2\arcsec   \\
30 Sep 2013 & 85965 & 56565.9351702(10) & 2       & LT+RISE   & V+R          & 14.8 & clear, seeing 1.8\arcsec \\
06 Nov 2013 & 86391 & 56602.9205819(16) & 2       & LT+RISE   & V+R          & 15.3 & thin clouds, seeing 1.8-2.5\arcsec\\
10 Nov 2013 & 86433 & 56606.5670216(45) & 3       & TNT+USPEC & g$^{\prime}$ & 15.1 & cloudy, seeing 1.5-2\arcsec        \\
13 Nov 2013 & 86467 & 56609.5189097(21) & 2       & TNT+USPEC & Schott KG5   & 15.2 & thin clouds, seeing 1.5\arcsec     \\
18 Jun 2014 & 88973 & 56827.0904134(16) & 5       & INT+WFC   & g            & 15.7 & clear, seeing 1.5\arcsec           \\
19 Jun 2014 & 88985 & 56828.1322517(37) & 5       & INT+WFC   & g            & 15.7 & clear, seeing 2-3\arcsec           \\
26 Jun 2014 & 89066 & 56835.1647131(24) & 2       & LT+RISE   & V+R          & 16.1 & clear, seeing 2\arcsec   \\
\hline \hline
\end{tabular}
\end{center}
\end{table*}

\begin{figure}
\begin{center}
\includegraphics[]{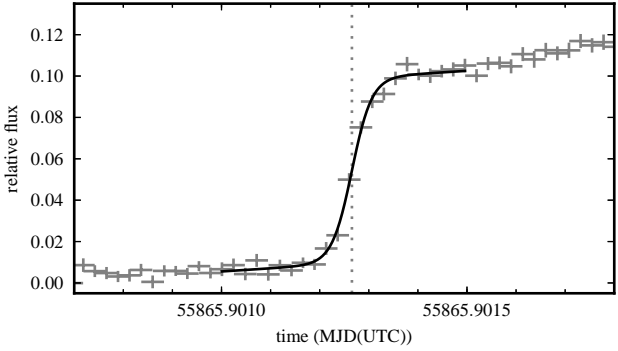}
\caption{Egress of the LT+RISE eclipse light curve of HU~Aqr, taken on 31 Oct 2011, with 2 second exposures. The solid black line shows the sigmoid+linear function that is fitted to the data. The vertical dotted line indicates the mid-egress time.}
\label{fig:sigmoid}
\end{center}
\end{figure}

As has been done for previously published times of HU~Aqr, we chose to measure mid-egress times as opposed to mid-eclipse times. This because the egress feature is relatively stable, even with the variable accretion rate, whereas the shape of the ingress feature varies significantly with the changing accretion rate and even differs from cycle to cycle. For all our new data we determined the time of mid-egress by a least-squares fit of a function composed of a sigmoid and a straight line,
\begin{equation}
y = \frac{a_1}{1 + e^{-a_2(x-a_3)}} + a_4 + a_5(x-a_3),
\end{equation}
where $x$ and $y$ are the time and flux measurements of the light curve, and $a_1$ to $a_5$ are coefficients of the fit. An example of one of our fits is shown in Fig.~\ref{fig:sigmoid}.

The straight line part allows us to fit the overall trend outside and during egress. This includes the egress of the white dwarf itself, which can have a significant contribution, especially when the system is in a low state. The sigmoid part of the function fits the sharp egress feature created by the egress of the accretion spot. To determine uncertainties, we have performed these fits in a Monte Carlo manner in which we perturb the value of the data points based on their uncertainties and we vary the number of included data points by a few at each edge, reducing any strong effects in the results caused by single datapoints. 

We converted all mid-egress times to barycentric dynamical time (TDB) in the form of modified Julian days and corrected to the barycentre of the Solar System, giving what we refer to as BMJD(TDB). The times are listed in Table~\ref{tab:times} and we used the ephemeris of \citet{Schwarz09} to calculate the corresponding cycle number $E$. Including the new times the best linear ephemeris is given by:
\begin{equation}
\mathrm{BMJD(TDB)} = 49102.42039316(1) + 0.0868203980(4) E.
\label{eq:ephemeris}
\end{equation}

\begin{figure}
\begin{center}
\includegraphics[]{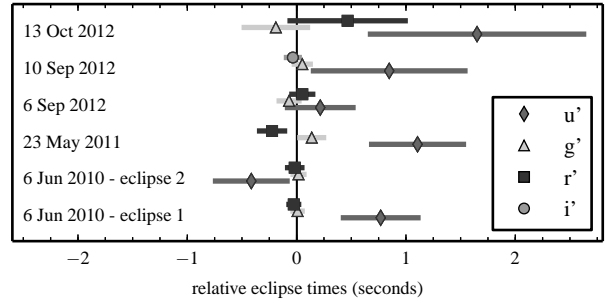}
\caption{Eclipse times relative to the weighted mean, with 1$\sigma$ errorbars, determined for data from the three ULTRACAM arms.}
\label{fig:relative_ultracam_times}
\end{center}
\end{figure}

For the ULTRACAM data we find that the times from the three individual arms agree well with each other, (Fig.~\ref{fig:relative_ultracam_times}), although the times from the blue arm are comparatively poor due to its lower time resolution (necessary to compensate for the lower flux in this band). The ULTRACAM times listed in Table~\ref{tab:times} are the error-weighted averages of the three individual times.

The agreement of the individual times and the absence of a particular ordering of the u$^{\prime}$, g$^{\prime}$, and r$^{\prime}$ or i$^{\prime}$ times around the weighted mean indicate that there is no significant correlation between the observed egress time and the wavelength at which the data were taken.

\section{Orbital period variations} \label{sect:orbitalperiodvariations}
\begin{figure*}
\begin{center}
\includegraphics[]{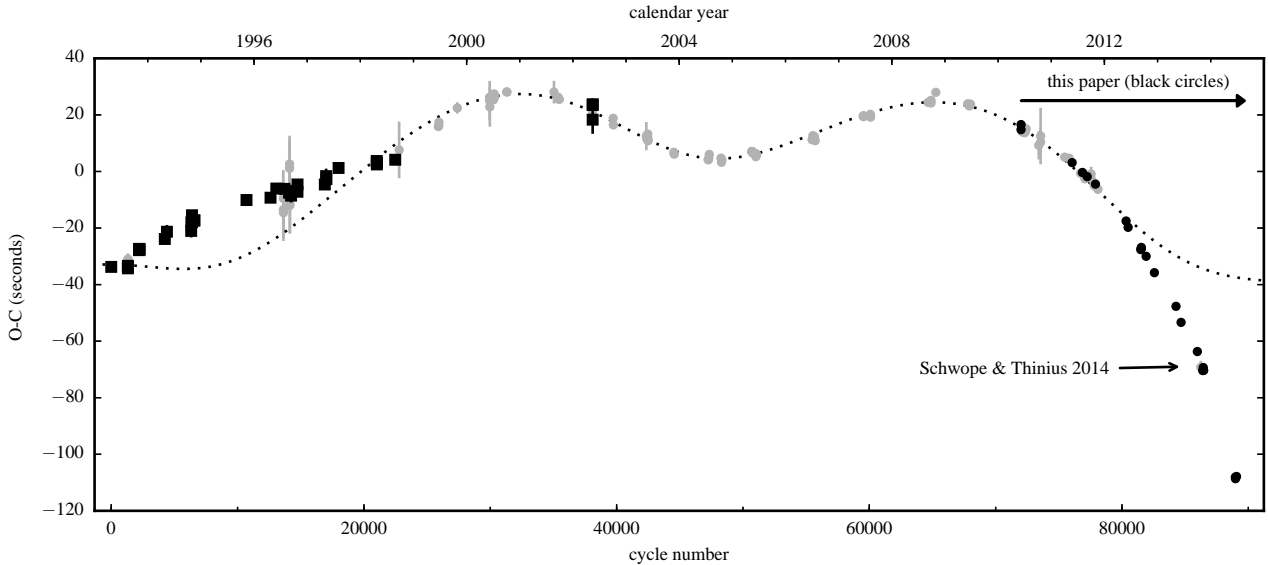}
\caption{O-C diagram for the mid-egress times of HU~Aqr with respect to the ephemeris in equation~\ref{eq:ephemeris}, including the new times presented in this paper (black dots) and literature times from optical (grey dots, excluding those from \citet{Qian11}) and non-optical data (black squares). The dotted line is the 1-planet model by \citet{Gozdziewski12}.}
\label{fig:oc_huaqr}
\end{center}
\end{figure*}

Fig.~\ref{fig:oc_huaqr} (referred to as an O-C diagram) shows the observed eclipse times minus times calculated assuming a constant orbital period. 

A mechanism that could explain observed O-C variations in white dwarf + M-dwarf binaries was proposed by \citet{Applegate92}. He suggested that magnetic cycles in the secondary star cause quasi-periodic variations in its gravitational quadrupole moment, which couple to the binary's orbit and cause semi-periodic variations in the orbital period. However, as orbital period variations were monitored more extensively and over longer periods of time, it became clear that, in HU~Aqr and other binaries, the variations can be larger than energetically possible with Applegate's mechanism. For the pre-CV NN~Ser, \citet{Brinkworth06} showed that the energy required for the observed variations was at least an order of magnitude larger than the energy available from the M-dwarf. Given that HU~Aqr has a similarly low-mass M-dwarf star, and the O-C variations are even more extreme, a similar discrepancy exists for this system \citep{Schwarz09}.

\subsection{Planetary companions to HU~Aqr}
For eclipsing white dwarf binaries that show O-C variations too large to be explained by Applegate's mechanism a number of models invoking circumbinary planets around close binaries have been suggested. These planets are generally in orbits with periods of years to decades, and would introduce periodic variations in the O-C eclipse times much like the ones observed. A comprehensive overview of the relevant binaries and models can be found in \citet{Zorotovic13}. Due to the long periods of the suspected circumbinary planets, and the often relatively short coverage of eclipse times, published models for planetary systems are only weakly constrained and are often proved incorrect when new eclipse times become available \citep{Gozdziewski12,Beuermann12,Parsons10}. Besides creating a model that fits the data, analysing the dynamical stability of the resulting system is a crucial step in determining the probability that circumbinary planets are present. Several published systems for which multiple planetary companions were proposed have turned out to be unstable on timescales as short as a few centuries \citep{Horner13,Wittenmyer13,Hinse12}, which makes their existence unlikely. Systems for which models invoke a single planetary companion are of course dynamically stable, but not necessarily more likely to exist. Only for the post-common envelope binary NN~Ser have planetary models (which include two planets) correctly predicted future eclipse times and shown long-term dynamical stability \citep{Marsh14,Beuermann13,Horner12a}, while at the same time both Applegate's mechanism \citep{Brinkworth06} and apsidal motion \citep{Parsons14} have been ruled out as the main cause of the eclipse timing variations. Despite the difficulties encountered, determining parameters of current-day planetary systems around evolved binary stars could provide a unique way to constrain uncertainties in close binary evolution, such as the common envelope phase \citep{PortegiesZwart13}, as well as answer questions related to planet formation and evolution.

HU~Aqr has been the topic of much discussion and speculation concerning possible planetary companions. A few years after the first egress times were published by \citet{Schwope01} and \citet{Schwarz09}, \citet{Qian11} published a model with two circumbinary planets, and a possible third planet on a much larger orbit. This model was proven to be dynamically unstable on very short timescales (10$^3$ - 10$^4$ years) by \citet{Horner11}, after which both \citet{Wittenmyer12} and \citet{Hinse12} reanalysed the data and found models with different parameters, which were nonetheless still dynamically unstable. \citet{Gozdziewski12} then suggested that there may be a significant correlation between eclipse times and the wavelength at which the relevant data is obtained, and proposed a single-planet model to explain the observed O-C variations seen in data taken in white light or V-band only, thereby excluding data that was taken in X-rays (ROSAT, XMM) and at UV wavelengths (EUVE, XMM OM-UVM2, HST/FOS) and all polarimetric data. They also excluded the outliers from \citet{Qian11}, which do not agree well with other data taken at similar times.

The O-C diagram as shown in Fig.~\ref{fig:oc_huaqr} includes all previously published eclipse egress times, except those from \citet{Qian11} which we exclude for the same reason as \citet{Gozdziewski12}. Also plotted is the 1-planet model that was proposed by \cite{Gozdziewski12}. Our new times and the time from \citet{Schwope14} depart dramatically from this model, and therefore we conclude that the proposed orbit is incorrect. The deviation of our new data also suggests that the times derived from satellite data (which \citet{Gozdziewski12} argued were unreliable) should be considered alongside optical data. This is also supported by the agreement between the optical and satellite times when taken at similar epochs. As mentioned before, from the separate u$^{\prime}$, g$^{\prime}$ and r$^{\prime}$/i$^{\prime}$ data we also do not see any evidence that mid-egress times are wavelength dependent.

For completeness we fitted the O-C times with three different planetary models, containing one, two and three eccentric planets respectively. We did not include a quadratic term, since such long-term behaviour can be mimicked by a distant planet, and chose to investigate a possible secular change of the orbital period separately in Sect.~\ref{sect:secularchange}. Unsurprisingly, all planets in our models are forced into highly eccentric orbits in order to fit the recent steep decline in the O-C times, leading us to believe that these system would be dynamically unstable. In addition, the models leave significant residuals. We conclude that the observed variability in eclipse egress times is not caused purely by the presence of a reasonable number of circumbinary planets.

\subsection{A secular change of the orbital period?} \label{sect:secularchange}
\begin{figure}
\begin{center}
\includegraphics[]{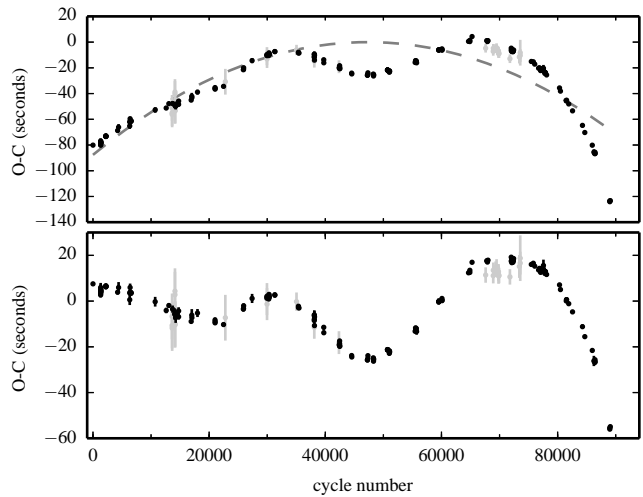}
\caption{O-C diagram of the mid-egress times of HU~Aqr including the best quadratic fit as the dashed grey line (top panel), and the residuals relative to that fit (bottom panel). O-C times with uncertainties exceeding 3~seconds have been greyed out for clarity, but all data were used to obtain the fit.}
\label{fig:oc_quad}
\end{center}
\end{figure}

It seems that the current set of O-C times cannot be explained simply by a model that introduces one or multiple planets, but given the recent steep decline in the O-C times, we now consider whether we are seeing the long-term evolution of the binary's orbital period. Using a quadratic model, shown in Fig.~\ref{fig:oc_quad}, we measure the rate of orbital period change as -5.2~$\cdot~10^{-12}$~s/s (= -4.5~$\cdot~10^{-13}$~days/cycle). 

With an orbital period of only 125 minutes, HU~Aqr is located just below the CV period gap \citep{Knigge06}, so that gravitational wave emission is expected to be the main cause of angular momentum loss. Using M$_{\mathrm{wd}}$~=~0.80~M$_{\odot}$, M$_{2}$~=~0.18~M$_{\odot}$ for the masses of the two stars \citep{Schwope11}, and the orbital period from equation~\ref{eq:ephemeris}, we calculated the period change due to gravitational wave emission to be $\dot{P}_{\mathrm{GW}}$ = -1.9~$\cdot~10^{-13}$~s/s, 27 times smaller than the result from our best fit to the O-C times. If the measured quadratic term in the O-C times represents an actual change in the binary's orbital period, this is not caused by gravitational wave emission alone. For binaries that lie below the period gap it is likely that some magnetic braking is still ongoing \citep{Knigge11}. If this occurs in short bursts of strong magnetic braking, rather than long-term steady magnetic braking, it might be possible to create large changes in the binary's orbital period on short timescales.

\subsection{Magnetic alignment of the accretion spot}
A last possibility we consider is that the accretion spot, the egress of which is the feature being timed, moves with respect to the line of centres between the two stars. This could happen either because of asynchronous rotation of the white dwarf \citep{Cropper88}, or because the spot itself migrates on the surface of the white dwarf \citep{Cropper89}. 

From eclipse data taken when HU~Aqr was in a low state we know that the ingress and egress of the white dwarf last for $\sim$~30~seconds \citep{Schwarz09}. Therefore the maximum libration of the spot on the surface of the white dwarf can generate O-C deviations with an amplitude of 15~seconds. For geometrical reasons, this would have to be accompanied by shifts in the time of maximum light from the accretion spot of $\sim$~0.25 orbital phases, an effect that has not been observed. Additionally, even with a large quadratic term removed from the original O-C times, an amplitude of 15~seconds is not large enough to explain the residuals, which still fluctuate by more than 20~seconds (Fig.~\ref{fig:oc_quad}). Furthermore, we find no correlation between the accretion state of the binary and the magnitude of the O-C deviations, in the original O-C diagram, nor in the residuals after removal of the best quadratic fit.

\section{Conclusions}
We have presented new eclipse observations across the optical spectrum of the eclipsing polar HU~Aqr while in high as well as low accretion states. From the egress feature of the accretion spot on the white dwarf we have determined the mid-egress times. Our ULTRACAM data shows that times from data taken using different SDSS filters agree well with each other. The new O-C times indicate that the eclipses are still occuring increasingly earlier than expected when using a linear ephemeris and a constant orbital period. They also confirm the result found by \citet{Schwope14} that the circumbinary planet proposed by \citet{Gozdziewski12} does not exist, nor can the entire set of egress times be well fitted by a model introducing one or multiple planets. 

Given the large amplitude of the observed O-C times, Applegate's mechanism can likely be excluded, and also asynchronous rotation of the white dwarf or movement of the accretion spot seem unlikely. Also a long-term orbital period change induced by gravitational wave emission or constant magnetic braking is not large enough to explain the observed O-C variations. Currently, our best guess is that more than one of these mechanisms act at the same time, working together to produce the dramatic eclipse time variations observed in this binary.

\section*{Acknowledgements}
We thank the referee (Robert Wittenmyer) for the prompt and positive feedback. TRM acknowledges financial support from STFC under grant number ST/L000733/1. SGP acknowledges financial support from FONDECYT in the form of grant number 3140585.

\bibliographystyle{mn_new}
\bibliography{bibliography}

\label{lastpage}
\end{document}